# The interfacial spin modulation of graphene on Fe(111)

*Jeongmin Hong[†][*], Han-Na Hwang[††], Alpha T. N'Diaye[†††], Jinghua Liang[¶], Gong Chen[#], Youngsin Park[※], Laishram T. Singh[††], Yong Gyun Jung[€], Ji-Hoon Yang[$], Jae-In Jeong, Andreas K. Schmid[#], Elke Arenholz[†††], Hongxin Yang[¶], Jeffrey Bokor[⊥], Chan-Cuk Hwang[††]*, Long You[†]**

[†]School of Optical and Electronic Information, Huazhong University of Science and Technology, Wuhan 430074, People's Republic of China.

[††]Pohang Accelerator Laboratory (PAL), Pohang University of Science and Technology, Pohang 37673, Republic of Korea.

[†††]Advanced Light Source, Lawrence Berkeley National Laboratory, Berkeley, CA 94720, USA.

[¶]Key Laboratory of Magnetic Materials and Devices, Ningbo Institute of Materials Technology and Engineering, Chinese Academy of Sciences, Ningbo 315201, People's Republic of China.

[#]The Molecular Foundry, Lawrence Berkeley National Laboratory, Berkeley, CA 94720, USA.

[※]School of Natural Science, UNIST, Ulsan 44919, Republic of Korea.

[€]Steel Solution Marketing Planning Group, POSCO, Seoul 06194, Republic of Korea.

$Research Institute of Industrial Science and Technology (RIST), Pohang 37673, Republic of Korea.

⊥EECS, University of California, Berkeley, California 94720, USA.

* E-mails: jehong@hust.edu.cn; lyou@hust.edu.cn; cchwang@postech.ac.kr

When Fe, which is a typical ferromagnet using *d*- or *f*-orbital states, is combined with 2D materials such as graphene, it offers many opportunities for spintronics. The origin of 2D magnetism is from magnetic insulating behaviors, which could result in magnetic excitations and also proximity effects. However, the phenomena were only observed at extremely low temperatures. Fe and graphene interfaces could control spin structures in which they show a unique atomic spin modulation and magnetic coupling through the interface. Another reason for covering graphene on Fe is to prevent oxidation under ambient conditions. We investigated the engineering of spin configurations by growing monolayer graphene on an Fe(111) single crystal surface and observed the presence of sharply branched, 3D tree-like domain structures. Magnetization by a sweeping magnetic field (m-H) revealed that the interface showed canted magnetization in the in-plane (IP) orientation. Moreover, graphene could completely prevent the oxidation of the Fe surface. The results indicate possible control of the spin structures at the atomic scale and the interface phenomena in the 2D structure. The study introduces a new approach for room temperature 2D magnetism.

**Introduction**

Recent discoveries on van der Waals 2D magnetism are based on well known magnetic insulators, such as $CrI_3$ and $Cr_2Ge_2Te_6$.[1—5] The effects of 2D magnetism are systematic and significant, but these insulators are limited regarding their actual applications due to the low temperature phenomena at 80 K. While graphene exhibits metallic conductivity in the in-plane (IP) orientation, it serves as an insulator up to a very high temperature above 500 K.[6] Continuous high-quality single-layer graphene is a paramagnet compared to other 2D magnetic crystals.[7,8] Possibly, the use of ferromagnet as tunneling barriers can offer unique opportunities.[9]

The magnetization of the Fe crystal along the (100) plane requires the least energy, whereas that along Fe(111) consumes the greatest amount of energy at approximately $3.5 \times 10^{-6}$ eV per atom.[10] As a strong ferromagnet, Fe with a 2D interface presents magnetic coupling due to the drastic change of the interface-induced magnetic anisotropy energy (MAE). For this reason, on the basis of the carbon-coated bcc-Fe(111) structure, it has been possible to engineer spin structures by manipulating the properties of the Fe(111) orientation. The spin structures align in a unique way that can form oriented domain structures, resulting in canted magnetization.

At the same time, the presence of graphene on Fe(111) prevents the oxidation of Fe and consequently preserves its unique properties, enabling the engineering of atomic spin structures. Once oxidized, the domain patterns realign in an energy-efficient manner, reducing domain walls as much as possible. Thus, the development of alternative forms of Fe-C-based materials is timely. To this end, the adsorption of well-defined graphene under ultra-high vacuum (UHV) conditions could be an excellent approach for engineering the magnetic properties of Fe. As previously reported, no impurities, even the smallest atoms, can penetrate graphene.[11]

Here, we report the observed properties of the interfaces between a graphene layer on single-crystalline Fe(111) after graphene growth via the dissociative adsorption of $C_2H_2$ under UHV conditions. Because graphene protects the underlying substrate, Fe(111) presents sharply branched 3D patterns due to the change in the MAE and step-edge energies. This combination of the materials, with its aforementioned advantages, could be critical for many practical applications in future electronics.

**Results and Discussion**

We synthesized a uniform structure of graphene on Fe(111)substrates. Compared with other synthesis methods, the dissociative adsorption of $C_2H_2$ is unique because it allows ultra-high-quality graphene to be grown under UHV conditions.[12] The details of the synthesis are described in the Methods section. Figure 1(A) shows a schematic model of the Fe(111) structure. The Fe(111) substrate is modeled as 6 layers of an Fe crystal. The length of the basic vector in the primitive surface unit cell of the Fe(111) film (a = 4.05A) is approximately 1.6 times larger than that of graphene (a = 2.46 A). The graphene on the Fe(111) structure is illustrated in Figure 1(B). Accordingly, we studied the graphene-(5×5)/Fe(111)-(3×3) system, in which the lattice mismatch is reduced to less than 2%. Therefore, only interface effects were considered in the calculation.

Figure 1(C) presents the band structures of bare Fe(111) (left panel) and graphene/Fe(111) (right panel) around the $\Gamma$ point of the Brillouin zone. The $\sigma_{2,3}$ and $\pi$ bands of graphene are clearly observed, and the binding energies of these states are observed at ~4.5 and ~8.5 eV below the Fermi energy, respectively. The band structures of pristine Fe and graphene/Fe(111) around the $K$ point are presented in Figure 1(D) (left and right panels, respectively). Figure 1(D) shows two clear $\pi$ bands near the Fermi level, indicating that high-quality graphene was synthesized on the Fe(111) surface. The interaction energy between graphene and Fe(111) is stronger that between graphene and other transition metals, such as Ni and Co.[13,14] Two different $\pi$ bands arise from two separately rotated graphene domains. The rotation of the band dispersion is a signature of dual domains. In addition, two parabolic $\sigma_1$ bands are observed, analogous to the two $\pi$ bands. The minima of the $\sigma_1$ bands around the $K$ point are measured to lie at ~10 eV below the Fermi energy. Multiple rotations of graphene have been similarly reported for graphene/Cu(111).[15]

Figure 1(E) shows the low-energy electron diffraction (LEED) pattern of graphene/Fe(111). The hexagonal lattice structure arises from the bcc-Fe(111). The other spots, A and B, indicate the formation of two rotated graphene structures on the Fe(111).[16,17] The rotation angle with respect to the hexagon of the Fe(111) is approximately ±15°. To better understand the rotation angle, Figure 1(F) shows the ARPES and LEED patterns together. The rotation angle is consistent with the angle-resolved photoemission spectroscopy (ARPES) results.

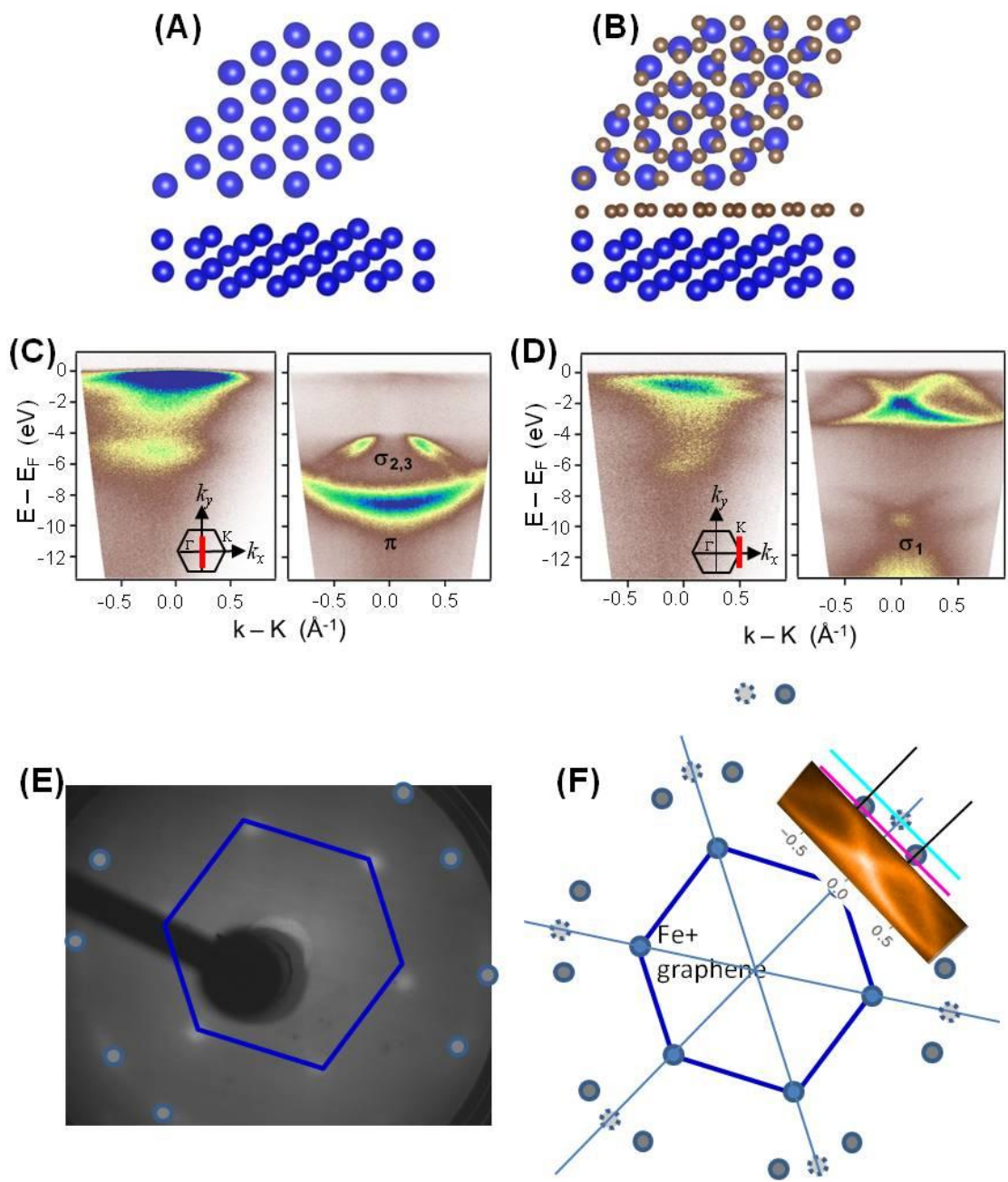

**Figure 1 | Schematic illustrations of the material structures, and the formation of the layers and the resulting band structures:** (A) Schematic of the bare Fe(111) structure. (B) Model of the structure after the formation of graphene on Fe(111). ARPES images of (C) bare Fe(111) and (D) graphene/Fe(111) around the *K* point. (E) The LEED pattern of

**graphene on Fe(111). (F) The pattern shows two separately rotated domains. The ARPES results also show dual $\pi$ bands, consistent with the LEED results.**

Scanning tunneling microscopy (STM) imaging under ambient conditions can identify pristine graphene structures.[18] Our analysis confirmed that the synthesized graphene was very smooth, as observed in the spectra shown in Figure 1. The inset image in Figure 2(A) shows branch-like lines that reflect the step-edge structure of the Fe under the graphene layer. This result confirms the occurrence of the Fe(111) orientation and the thermal effects during the synthesis. The topographical image originates from the Fe-graphene interface.

The occurrence of interface dynamics during graphene growth can be inferred. The graphene forms two main domains oriented at approximately ±15° with respect to the substrate, and effects resulting from the adsorption energy difference and thermal annealing of the Fe crystal structure are observed. Kinetic roughening could drive the random growth at the surfaces and interfaces in a certain way. The origin of this phenomenon is attributed to the domain rotation and kinetic roughening of graphene and Fe.[19,20]

Ultra-high-sensitivity magnetic force microscopy (MFM) images collected under ambient conditions show unique domain structures (Figure 2(B)).[21–23] The magnetic domain patterns show multiple sharply branched tree-like structures. Since several domain walls are not thermodynamically favorable, the reduction in the domains stabilizes with a low-energy structure (more favorable). As reported in the literature, monolayer graphene does not exhibit ordered magnetic properties; instead, it prevents the oxidation of the Fe substrates. Other types of Fe and Fe-based systems show similar patterns,[24,25] but graphene on Fe completely prevents other changes in domain patterns, such as oxidation.

To further investigate the magnetic orientation, we performed a spin-polarized low-energy electron microscopy (SPLEEM) experiment under UHV conditions. Since the depth of penetration is only 7 nm, it is a good candidate to probe interface effects. In SPLEEM, electrons are measured before and after they pass through a magnetic field. The results reveal unique domain patterns in a system without requiring the exposure of the sample to a magnetic field. SPLEEM produces clear images of the domain structures of Fe(111) with different magnetic orientations. The color scale in Figure 2(C) indicates the magnetic orientations from 0 to 360 degrees in the image. These types of unique domain patterns could not be observed in pristine Fe(111), even when argon annealing and sputtering were performed under ambient and UHV conditions. These unique domain patterns, which arise from the mixed states of step edge energy and MAE in this direction, are extremely small compared with the other orientations, and they can form randomly oriented, sharply branched domains.

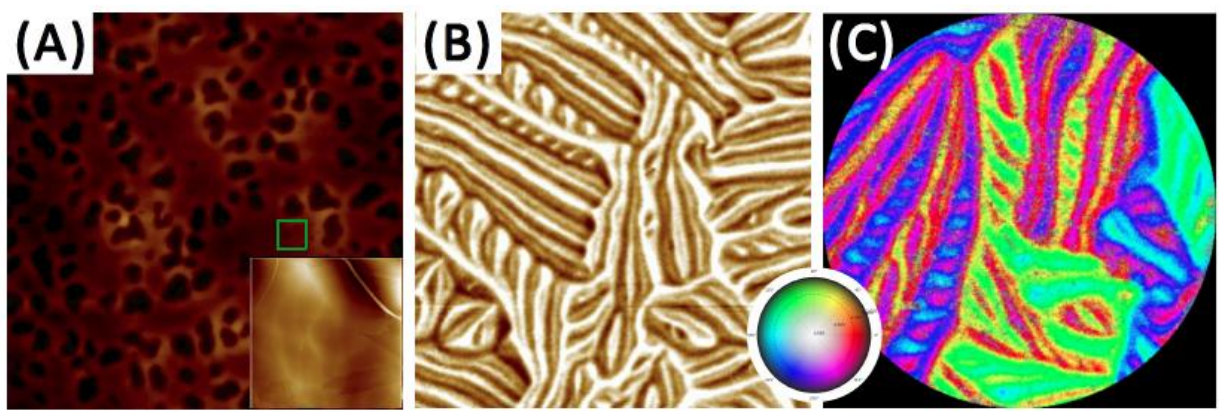

**Figure 2 | AFM, STM, MFM, and SPLEEM measurements. (A) Atomic force microscopy (AFM) image and (B) the resultant MFM images of the same structures under ambient**

conditions. The image dimensions are 5 × 5 $\mu m^2$. The inset shows an ambient STM image of monolayer graphene on single-crystalline Fe(111) (200 nm × 200 nm image). (C) A micrograph of the domain structures acquired via SPLEEM. The domain patterns show randomly oriented, sharply branched domain structures. The field of view (FOV) is 8 μm. The penetration depth of the beam is less than 5 nm.

We swept a magnetic field with respect to the magnetization at the Fe $L_{3,2}$-edge, as shown in Figure 3(A). The hysteresis loops indicate completely canted magnetization within the range of ±150 mT. The remanence within this range is zero. The plot is linear in the out-of-plane (OOP) component, indicating a hard axis. The interfaces are magnetically coupled to each other with zero remanence in this range, and for this reason, the unique domain patterns are presented. In contrast, the OOP components only show a hard axis, very similar to the behavior of bulk Fe(111).

As shown in Figure 3(B), the X-ray absorption spectroscopy (XAS) and X-ray magnetic circular dichroism (XMCD) spectra of the Fe $L_{3,2}$ edge present typical metallic behaviors without peak splitting or any signs of oxidation, although the system was exposed to air. The monolayer graphene membrane served as a coating to prevent any strong reaction or change in the interface anisotropy of the Fe(111). For bcc-Fe, $K_1$ is positive, and Fe(100) is the favorable direction. The highest density of atoms is in the (111) direction; consequently, the (111) axis is the hard axis, while (110) and (100) are the easy axes. The anti-parallel magnetization directions are crystallographically equivalent, providing three distinct easy directions for positive-$K_1$ systems.

A first-principles calculation was performed and the results showed that the MAE is reduced after the growth of the graphene interfaces. The topmost monolayer of Fe is coupled to graphene, changing the MAE to the other direction. The calculation procedure is detailed in the Methods section. The canted magnetization from the m-H loops is correlated with the randomly oriented tree-like domain patterns, as shown in Figure 2. Figure 3(C) shows the main results and the non-self-consistent total energy of the system was determined for the IP and OOP orientations of the magnetic moments. Couplings between monolayer graphene and transition metals due to proximity effects have been reported in the literature.[26,27]

Furthermore, the step-edge energy underneath the graphene layer induces atomic modulation and enhances the corrugation of the Fe surface during the annealing process. The surface morphology changes slightly due to the lattice mismatch. The Fe system changes to show unique domain patterns with canted magnetization in the IP direction. This modification allows closed 90° domains and unique domain structures in the Fe(111) orientation.

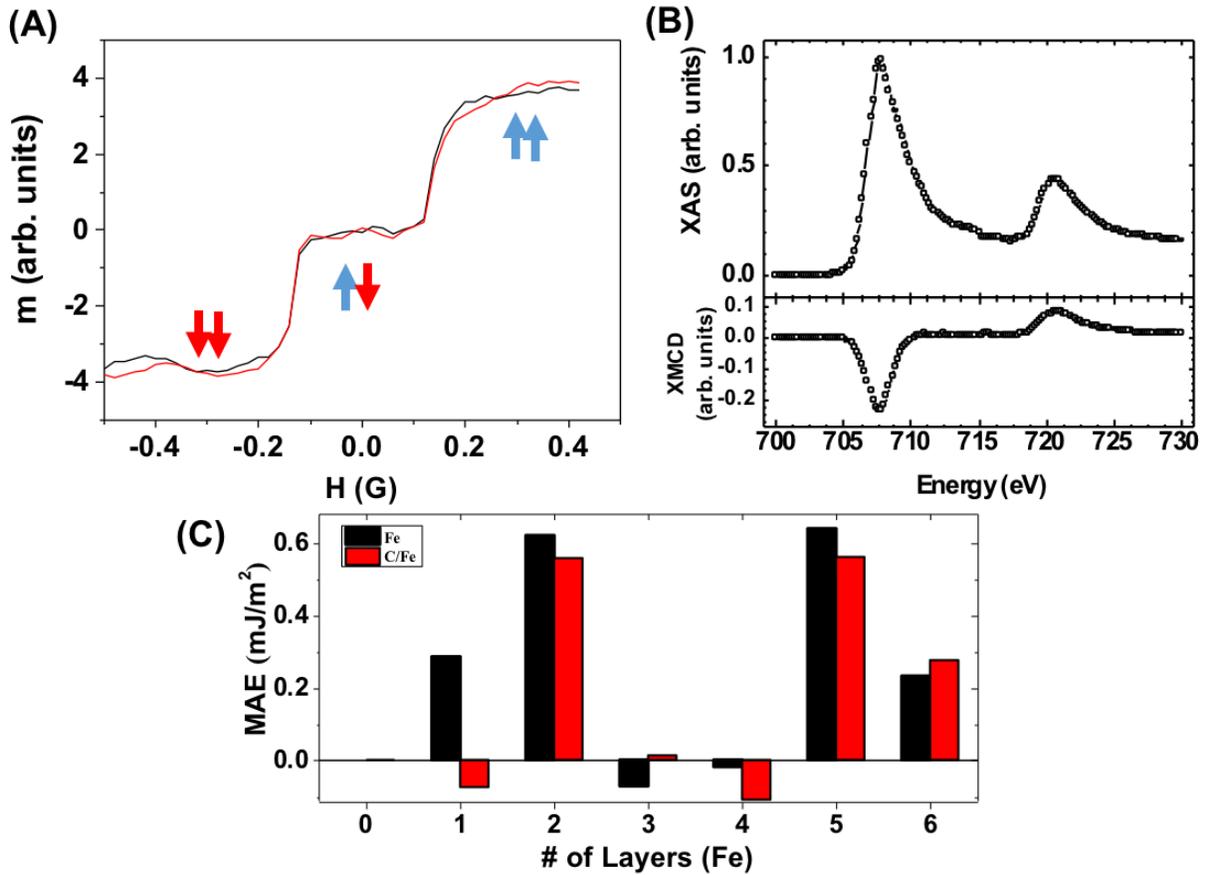

**Figure 3 | The M-H loops of the structure and the calculated and experimental X-ray spectra. (A) M-H loops of the structure in the IP orientation. (B) XAS and XMCD spectra of Fe(111) after graphene growth. The penetration depth of the X-ray is less than 7 nm. (C) First-principles calculation of the interface-induced MAE of graphene on the Fe(111) structure.**

The pre-cleaning of Fe crystals before synthesis is a critical step. The upper image in Figure 4(A) shows the wide-scan spectra of the bare Fe(111). Repeated sputtering and annealing rendered the Fe substrate ultra-clean, resulting in a spectrum with a relatively small C *1s* peak and a strong Fe *3p* peak near 50 eV. As shown in Figure 4(A), after graphene formation (lower panel), the C *1s* peak increases in intensity, while the Fe *3p* peak is significantly weaker. These results indicate that the Fe surface is covered with carbon atoms, and this dependence is regarded as a hallmark of monolayer deposition. A detailed view of this spectrum in a narrow range around the C *1s* peak is shown in the inset of the lower panel. The full width at half maximum (FWHM) of the C *1s* peak is approximately 0.58 eV, indicating the formation of high-quality graphene. This peak is substantially narrower than that of the CVD-grown graphene.[15]

Moreover, Fe readily undergoes oxidation and corrosion, which strongly limits its use in many applications, such as computer electronics.[28] Figure 4(B) shows the photoemission spectroscopy (PES) spectra of the clean and graphene-coated Fe surfaces with increasing $O_2$ exposure. The spectrum of the clean Fe surface (upper) shows a sharp Fe *3p* peak at 52.5 eV, and no other peaks are observed. As shown in the lower panel of Figure 4(B), however, when the surface is exposed to oxygen, the O *2s* peak is observed at ~21.95 eV, and the absorption of oxygen on the Fe surface substantially alters the valence band spectrum. The intensity of the O *2s* peak is saturated when less than 100 L of oxygen is applied. The Fe *3p* core-level peak also considerably changes due to the formation of bonds between oxygen and Fe atoms.

By contrast, when the graphene-covered surface is exposed to oxygen, the Fe *3p* peak remains nearly unchanged, and the O *2s* peak is not visible even up to 3000 L. These results indicate that oxygen atoms cannot penetrate the graphene and that they do not combine with Fe atoms. The graphene completely protects the Fe surface from oxidation. This protective effect was also confirmed by the ferroxyl test, as described in the Supplementary Materials section.

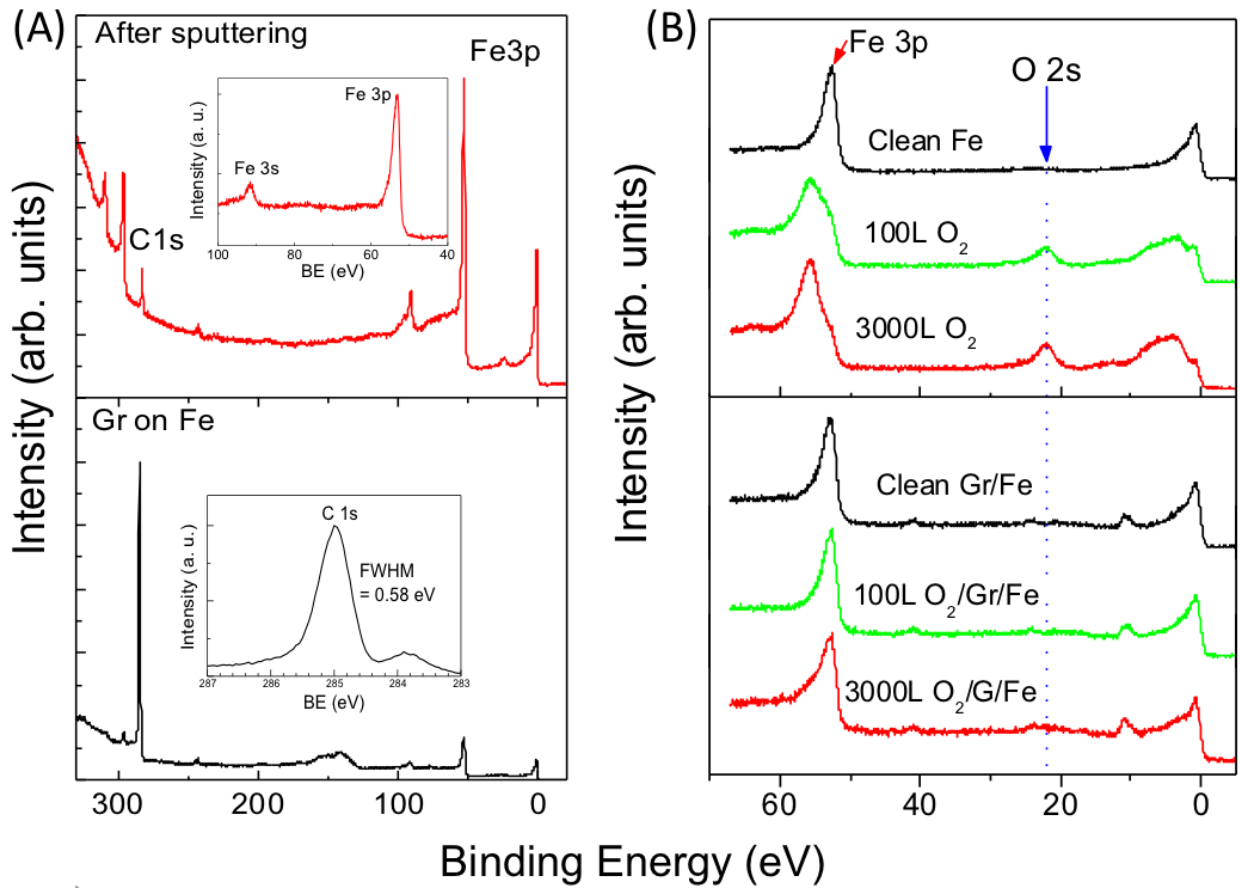

**Figure 4 | (A) PES spectra of pristine (upper) and graphene-coated (lower) iron surfaces. The inset of the lower panel shows the C *1s* peak in detail. (B) The PES results for pristine and graphene-coated Fe(111) surfaces after oxygen exposure up to 3000 L.**

**Conclusion**

Monolayer graphene on single-crystalline Fe(111) was carefully studied. We concluded that high-quality monolayer graphene growth is possible through dissociative adsorption under UHV conditions. Because of the presence of graphene, the iron substrate can retain its original properties because it is prevented from reacting with the oxygen in the air. Due to the unique properties of the Fe(111) structures, large numbers of randomly oriented domain walls are observed. In a magnetic field range of ±150 mT, the IP orientation is completely canted (coupled with the first layer of Fe atoms). If the structures are patterned along the domains, then high-quality spin devices can be simply fabricated by growing graphene on top of the surface.[28]

Sharply branched, tree-like domains form as a result of interfacial atomic spin reorientation. The evolution of these domains is caused by changes in the MAE; thus, it is possible to construct spin devices by manipulating the domain structures of Fe(111). In summary, graphene modulates the domain patterns of Fe(111). Thus, selectively patterned structures can be exploited in naturally grown 2D spin devices.

**Materials and Methods**

**Fabrication of monolayer graphene onto a single-crystalline Fe(111)** A clean Fe(111) surface was prepared by applying multiple successive cycles of $Ar^+$ sputtering at an energy of 1 keV for 30 min over ~24 h and subsequent annealing for 2–5 min at approximately 1000 K in a UHV chamber; this procedure was performed until carbon was minimized and no peaks corresponding to contaminants such as S, $NO_2$, or $O_2$ were observed by PES. After additional confirmation of the surface purity with LEED and ARPES, a graphene layer was grown on the Fe(111) substrate at ~1000 K in a $C_2H_2$ atmosphere of $1 \times 10^{-6}$ to $5 \times 10^{-6}$ torr.

**ARPES and XPS measurements** The ARPES studies were performed at beamline 10D of the Pohang Accelerator Laboratory (PAL), which is equipped with a Scienta R4000 analyzer that provides an overall energy resolution of ~50 meV at ~34 eV under a pressure of $1.2 \times 10^{-10}$ torr. The binding energy was calibrated by measuring the Au Fermi energy and the Au *4f* core-level spectrum. The data were collected at room temperature. The electronic band structures were taken from the clean Fe(111) surface and from the graphene grown on Fe(111) at the $\Gamma$ and $K$ points along the $k_y$ direction in the momentum space, as shown in the insets of Figures 1(C) and (D). A photon energy of 420 eV was used for spectrum collection.

**Ferroxyl spray test** The ferroxyl test was conducted while monitoring the appearance of the Fe surface. The test solution was sprayed onto the surfaces of the bare Fe and the graphene-coated structure.

**SPM and STM** Scanning probe microscopy (SPM) and MFM were performed via 30 nm dynamic mode scanning. STM was used for high-quality topographic measurements under ambient conditions.

**SPLEEM** Real-space images were acquired using three orthogonal electron beam spin-alignments such that the magnetic contrast along three orthogonal directions corresponds to the OOP magnetization direction and two orthogonal IP axes. SPLEEM images map the magnetization of the sample in the sense that the intensity in each pixel represents the dot product of the spin polarization vector **P** of the illumination beam and the magnetization vector **M** with a lateral resolution on the order of 10 nm.

**First-principles calculation** In the framework of density function theory, our first-principles calculation was performed using the Vienna ab initio simulation package (VASP)[29] with the Perdew-Burke-Ernzerhof generalized gradient approximation (GGA-PBE).[30] The electron-ion interaction is described by the projected augmented wave (PAW) potentials.[31] In all calculations, a kinetic energy cutoff of 520 eV and a $\Gamma$-centered 6×6×1 $K$-point mesh for the first Brillouin zone integration are employed. The MAE is calculated in three steps. First, structural relaxation is performed until the forces on each atom are smaller than 0.001 eV/Å to determine the ground state. Next, the Kohn-Sham equations are solved with no spin-orbit coupling (SOC) taken into account to determine the charge distribution of the system's ground state. Finally, the SOC is included, and the non-self-consistent total energy of the system is determined when the orientations of the magnetic moments are set both IP and OOP.

AUTHOR INFORMATION

**Corresponding Author**

JH (jehong@hust.edu.cn); LY (lyou@hust.edu.cn); C-C H (cchwang@postech.ac.kr)

**Author Contributions**

The manuscript was written through contributions of all the authors. All authors have approved to the final version of the manuscript.

**ACKNOWLEDGMENT**

This work was supported by the National Natural Science Foundation of China under Award number 61674062. The work was also supported by the U. S. Department of Energy, Office of Basic Energy Sciences, Division of Materials Sciences and Engineering under Contract No. DE-AC02-05CH11231. The authors acknowledge the financial support from the National Science Foundation (NSF) under Award number 0939514. This work was also supported by the National Research Foundation of Korea supported by the Korean Government (Ministry of Science and ICT) #2011-0030787, #2017R1A2B2003928, and #2017M3A7B4049173.

**Supplementary Materials**
   Materials and Methods

I. **Supporting Figures**

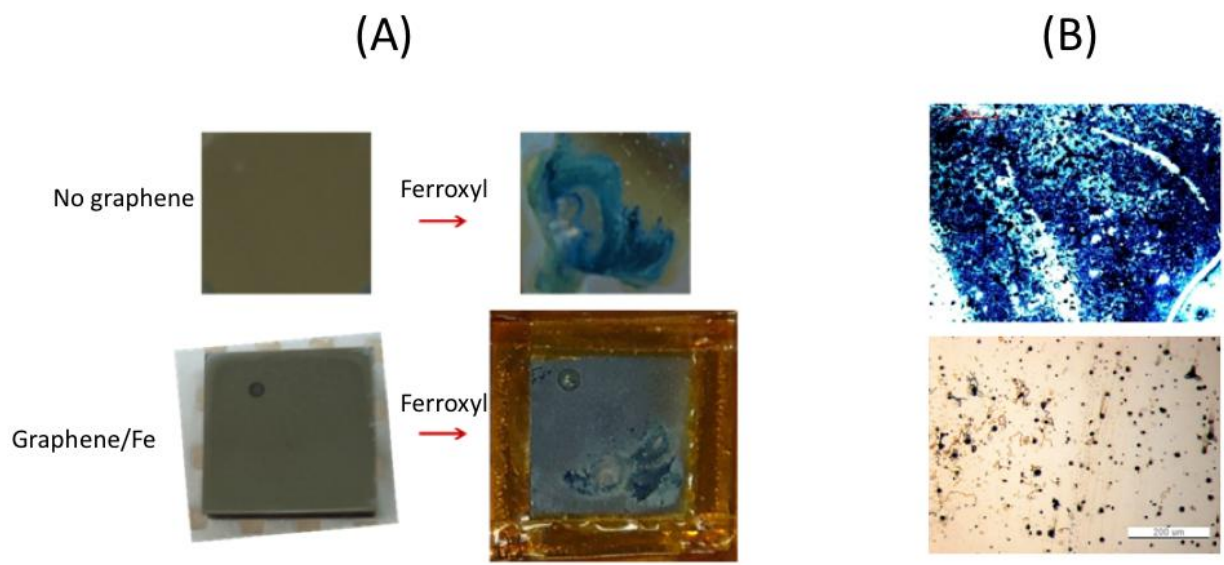

S1. (A) Photographs of bare (up) and graphene coated (bottom) Fe(111) single crystals before (left) and after (right) the ferroxyl test. (B) Optical microscopy images of bare (up) and graphene coated (bottom) Fe(111) single crystals after the ferroxyl test.

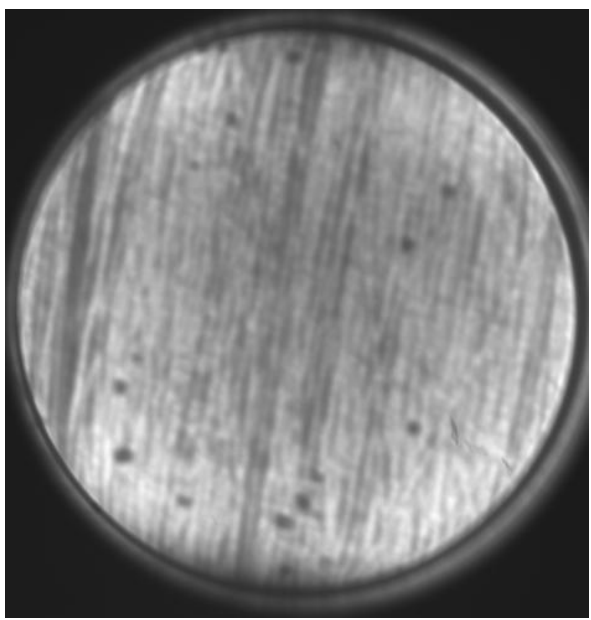

S2. LEEM image of monolayer graphene on single crystalline Fe(111). The FOV is 8 μm.